\documentstyle[12pt]{article}

\newcommand\bg{\begin{eqnarray}}
\newcommand\ed{\end{eqnarray}}
\newcommand\bgn{\begin{eqnarray*}}
\newcommand\edn{\end{eqnarray*}}

\def\del{\partial}

\begin{document}
\ 
\rightline{PUPT-1645\qquad}
\rightline{August 1996}
\vskip1in
\centerline{\bf MULTIPLE VACUA AND BOUNDARY CONDITIONS}
\centerline{\bf OF SCHWINGER-DYSON EQUATIONS}

\leftline{}
\leftline{}
\centerline{Zachary Guralnik}
\centerline{Department of Physics}
\centerline{Princeton University}
\centerline{Princeton, NJ 08544}
\centerline{zack@puhep1.princeton.edu}

\leftline{}
\leftline{}
\leftline{}

\centerline {\bf abstract} 
\leftline{}
\leftline{}
	We discuss the relationship between the boundary conditions of 
the Schwinger-Dyson  equations and the phase diagram of a bosonic field
theory or matrix model.  In the thermodynamic limit,  many boundary
conditions lead to the same solution,  while other boundary conditions
have no such limit.  The list of boundary conditions for which a 
thermodynamic limit exists depends on the parameters of the theory. 
The boundary conditions of a physical solution  may be quite exotic,  
corresponding to path integration over various inequivalent complex 
contours.

\leftline{}
\leftline{}
\leftline{}
\leftline{}
\leftline{}
\leftline{}
\leftline{}
\leftline{}
\leftline{}
\noindent{\it Talk given at
3rd AUP Workshop on QCD: Collisions, Confinement, and Chaos, Paris, 
France, 3-8
Jun 1996. }

\pagebreak
\noindent{\bf Introduction}\\[12pt]
\indent When attempting to solve a field theory or matrix model by 
using Schwinger--Dyson 
equations,  one must address the problem that these equations
do not possess a unique solution.   This problem came to our attention 
when trying to numerically solve
the equations for certain lattice field theories using a 
technique known as 
Source Galerkin \cite{firstpaper}, which is discussed elsewhere in 
these proceedings.
It was found that to make the numerical method stable,  one had to find 
some way of
selecting the right boundary conditions.  This led us to the more 
general question
of how the boundary conditions are related to the phase diagram of a field 
theory or
a matrix model.  We shall summarize our work on this relation in this paper.
Most of the details will appear elsewhere \cite{ourstuff}.   
 
The naive resolution of the problem of how to select the boundary 
conditions 
is to simply
pick the solution which corresponds to path integration over real fields.  
However, within certain phases of many theories,  it can be shown that 
this  
solution is actually not the physical one.  
Furthermore in theories with actions unbounded below, such as certain 
matrix models and
Euclidean Einstein gravity, the integral over real 
fields is not even convergent.  This forces the 
consideration of  ``exotic'' solutions  
of the Schwinger-Dyson equations,  which have representations involving 
sums of integrals of the fields over various inequivalent contours in 
the complex plane.
For theories with a local order parameter,  symmetry breaking solutions are 
naturally generated by choosing a symmetry breaking set of contours.  
In the conventional approach
to obtaining the broken phase, the real contour is chosen but a 
small symmetry 
breaking term is added to the action.
This term is removed only after taking a thermodynamic limit, in which 
the number
of degrees of freedom becomes infinite.  In fact one can also obtain
the broken phase by choosing a symmetry breaking 
boundary condition (contour)  and then 
taking the thermodynamic limit directly.   This is a simple example 
showing that
the exotic solutions are not necessarily unphysical.  
Even for actions which are bounded below, it is not always 
possible to obtain all physical solutions from
the integral over real fields.  There are explicit matrix model examples
of this.
We conjecture that 
exotic solutions may also be physical for 
theories with a nonlocal order parameter,  though this has
yet to be demonstrated.  

The difficulty in choosing the correct boundary conditions lies in the 
fact that 
there are so many of them.   Furthermore since the Schwinger-Dyson 
equations 
satisfied by
the partition function are linear,  there naively appears to be a 
continuum of mixed phases,
which does not make physical sense.  Most of this problem is 
resolved by taking the thermodynamic limit.  In this limit the 
solutions associated with many boundary conditions 
coalesce,  while other boundary conditions do not lead to solutions 
with a thermodynamic limit.
In general one is left with a discrete set of solutions,  
although sometimes a continuum of distinct solutions survives, 
labeled by a countable number of theta parameters.
The set of solutions which survives in 
the thermodynamic limit may include both the physical vacua,  and 
false vacua with
complex free energy,  as well as other unphysical solutions which must 
be discarded by hand.
However, most of the task of discarding solutions is done 
automatically in the thermodynamic limit. It can be shown that
the set of boundary conditions for which this limit exists 
varies 
along paths in the space of coupling constants.  This set may vary smoothly 
or it may
be discontinuous at certain points along the path.  We have found 
it easiest to study
the behavior of this set in the context of one matrix models,  while field 
theories have
proven less tractable.   However we suspect that our conclusions are 
quite general.
One important conclusion is that the set of boundary conditions with a 
thermodynamic limit
changes discontinuously as one crosses a phase boundary.
\leftline{}\\[12pt]
\noindent{\bf Boundary Conditions In Zero Dimension}
\leftline{}
\leftline{}
\indent To illustrate the multiple solutions of the 
Schwinger-Dyson equations we begin with 
the simple example of a zero dimensional theory 
with a polynomial action, 
$S(\phi) = {1\over n}g_n \phi^n$.  The Generating functional of 
disconnected green's 
functions $Z(J)$ satisfies the Schwinger--Dyson equation 
\bg
(g_n {\del\over \del J^{n-1} } - J)Z(J) = 0
\label{poln}
\ed
The order of this equation is determined by the highest order term in the 
polynomial action.  If the highest order term 
is $\phi^k$ then there is an $k-1$ parameter space
of solutions.  One of these parameters is just the overall normalization 
of $Z$, so
there is a $k-2$ parameter space of solutions with distinct green's 
functions.  
An integral representation
of the solutions is
\bg
Z(J) = \int_{\Gamma} d \phi  e^{-S(\phi) + J \phi}
\label{complexcont}
\ed
It is easy to show that this satisfies the Schwinger-Dyson equation 
provided that
\bg
\left. e^{-S(\phi) + J\phi} \right|_{\partial \Gamma} =0
\label{bterm}
\ed
This condition is satisfied if $Re S(\phi) \rightarrow +\infty$ 
asymptotically on the contour
$\Gamma$.  For the polynomial action the condition becomes 
$Re g_k\phi^k \rightarrow +\infty$.  Therefore there are k wedge 
shaped domains in
the complex plane in which the contour can run off to infinity.  
There are $k-1$ 
independent contours satisfying (3),  as expected from the order of the 
Schwinger-Dyson
equation.   In general the behavior of the action for large 
fields 
controls the solution set,  even for many degrees of freedom.  The 
condition (3) can be used to construct this set even when the action
is non polynomial and the order of the Schwinger-Dyson equations is 
unclear. 
For example the action $S=\beta cos \phi$,  (one plaquette QED), can be 
shown 
by this method to 
yield a two parameter class of solutions for $Z$.
A basis set of solutions is given by the contours 
\bg
\Gamma_1 = [-i\infty, +i\infty]
\label{kal}
\ed
and
\bg
\Gamma_2 =[-i\infty, 0] + [0, 2\pi] + [2\pi, 2\pi + i\infty]
\label{cal}
\ed
The difference between these two solutions is the usual solution in which 
the 
contour runs from $0$ to $2\pi$.
Note that in general the exotic solutions for a gauge theory correspond to 
a 
complexification of the gauge group.
In this simple case it is 
possible to verify the counting of solutions by coupling sources
$J$ and $\bar J$ to the loop variables $e^{i\phi}$ and $e^{-i\phi}$. 
The Schwinger-Dyson equation is then,
\bg
[\beta(\del_J - \del_{\bar J} ) - 
2(J\del_J - {\bar J}\del_{\bar J} )]Z(J,{\bar J}) = 0
\label{plaq}
\ed
which when combined with the constraint,
\bg
{\del\over\del J}{\del\over\del \bar J}Z = Z
\label{pha}
\ed
yields a two parameter class of solutions for $Z$.
\\[24pt]
\leftline{}
\noindent{\bf Boundary Conditions in the General Case}\\[12pt]
\leftline{}
\indent Let us consider the solution set for a lattice field theory. 
For theories in which the large field behavior of the action is 
dominated by 
independently variable local 
terms, such as $g_k \phi^k(x)$,  the construction of the space of 
solutions is a simple 
generalization of the zero
dimensional case. One  simply 
chooses one of the zero dimensional contours
for each field at each lattice site.  An arbitrary solution is 
obtained by summing
solutions with a definite set of contours at each site. 
The solution set is then somewhat reduced by 
imposing the lattice symmetries,  but is still very large.  
In a one matrix model,  the lattice site label is replaced by an 
eigenvalue label, 
and since the interaction between eigenvalues is only logarithmic,  the 
highest order
term in the potential determines the allowable contours for each 
eigenvalue.   
For theories in which the large field behavior of the action is not 
dominated by independently variable local
terms,  the construction of the solution set is somewhat more complicated.  
This appears to
be the situation for many lattice theories with a non-local order parameter.

For a generic action with a finite number of degrees of freedom, 
the number of independent solutions of the Schwinger-Dyson equations is
exactly equal to the number of classical solutions,  complex ones 
included.  
The exception to this rule occurs
when the action has flat directions, or extrema with vanishing second 
derivative,
in which case every term in the perturbative expansion about the classical 
solution diverges.
For example in zero dimensions,  the potential $V= g\phi^4$  has only one 
classical solution,  but there are three independent solutions of the 
Dyson 
Schwinger equations.  Assuming that we are not considering such an 
exceptional case, 
the borel re-summation of the perturbation series about any classical 
solution
can be shown to  yield
one or more exact solutions of the Schwinger-Dyson equations with some 
exotic 
integral representation.  One can see to which contours these solutions 
must correspond 
by taking the  weak coupling limit.  The contours,  deformed so that 
$e^{-S}$ has constant
phase,  must avoid 
all the classical solutions of equal or lower action than the one whose 
perturbative 
expansion is being borel re-summed.  There are often many choices of such 
contours.  In terms 
of the inverse borel transform,
\bg
Z_i(g)= \int_{t=0}^{Re t = + \infty} e^{-t/g}B_i(t) 
\label{brl}
\ed
these choices 
correspond to various various ways of avoiding positive real singularities 
in the 
Borel variable $t$. In the above equation the index $i$ labels a classical 
solution.
Thus the complete set of solutions of the classical equations
yields an over-complete set of solutions of the Schwinger-Dyson equations. 
An arbitrary solution is obtained by taking linear combinations of these
 solutions,  

Thus away from the thermodynamic limit,  there are a continuum of phases, 
resembling 
theta vacua.  The resemblance to conventional theta vacua is actually 
quite strong.  
In the conventional approach to 
theta vacua, such as those in a Yang Mills theory,  a particular theta 
vacuum is selected by 
adding a surface term to the action without any recourse to exotic 
contours.  
This alters the Schwinger Dyson equations only at the
space time boundary,  and therefore in the infinite volume limit the role 
of the surface term 
is to set a 
boundary condition for these equations.  
The surface term puts a different weight on the 
contribution to the generating function coming from fluctuations about 
different 
classical sectors.
As we have seen above,  this is equivalent to putting  different weights 
on different 
complex contours,  but without any modification of the action.  Note 
however that it 
is not always easy or even possible to obtain all solutions
of physical interest from the real integral, assuming it is convergent, 
by modifying  the action with a surface term,  
or by small perturbations of the action which are removed after taking the 
thermodynamic limit.
It is generally impossible to obtain false vacuum solutions in this way.  
Furthermore, there are 
examples of symmetry breaking solutions of matrix models
which appear to be impossible to obtain
from the real contour by any perturbation of the action [4].  
The complete set of independent complex contours however
accounts for all the phases of the theory.
 
To complete the classification of these phases one needs 
a rule for identifying solutions in the
same phase at different values of the coupling constants. 
More precisely, one needs some 
set of first order differential equations in the coupling constants,  
which the 
partition function in
a given phase should 
satisfy.  A natural choice is given by the Schwinger action principle,  
which may be 
stated as,
\bg
({\del\over\del g}-<{\del S \over \del g}>) Z=0 
\label{swac}
\ed
or, for the example of the zero dimensional polynomial action, as
\bg
({\del\over\del g_n}-{1\over n}{\del\over\del J^n})Z=0
\label{wic}
\ed
The action principle means that some set of contours and weights of 
contours 
associated with a solution is held fixed as one makes an infinitesimal 
change in the
coupling constants.  Note that 
there are classes  of contours which lead to the same solution,
either by being mutually deformable,  or by leading to solutions which 
coalesce in
the thermodynamic limit.  However as one changes the couplings, these 
classes also change.
Contours which were convergent may become non-convergent,  and contours 
for which a 
thermodynamic limit existed may no longer have such a limit. 
Thus, the action principle does not allow one to fix the contours and 
weights globally in the space of coupling constants.  
There may be branch cuts 
in certain coupling constants, with branch points as phase boundaries.  
For instance in the
zero dimensional $g\phi^4$ theory, if one rotates the phase of the 
coupling constant by 
$2\pi$  in accordance with the action principle,  the contour of 
integration rotates by 
${\pi\over 2}$ to maintain convergence, yielding an inequivalent solution. 
Thus the action principle fixes the phase only 
locally in the space of coupling constants. 

Note that one could have chosen another set of first order differential 
equations
in the couplings instead of the action principle. 
However the Schwinger action operators which annihilate the generating 
function commute 
with the Schwinger-Dyson operators which annihilate the generating 
function.  
If a continuum
of solutions to the Schwinger-Dyson equations reduce to a discrete set 
in the 
thermodynamic
limit,  then in general the action principle is automatically induced.  
For instance
matrix
model solutions of the Schwinger-Dyson equation  automatically satisfy 
the action
principle in the planar limit.
If theta parameters survive in a thermodynamic limit,  then the action 
principle is
not induced,  but may certainly be consistently imposed.
\\[24pt]
\leftline{}
\noindent{\bf Vacuum Selection}\\[12pt]
\leftline{}
\indent 
Having classified the solutions far from the thermodynamic 
limit,  one 
needs some way of reducing the solution set.  It is tempting to invoke 
such requirements
as reality and positivity.  Reality does not throw out all exotic solutions.
If the coupling constants 
of the theory are real,  one can always take linear combinations of exotic 
solutions 
with complex contours such
that all Green's functions become real.  Thus reality is not so strong a 
constraint and also
throws away false vacuum solutions which may be of physical interest.  
In seeking the true vacuum for for an action bounded below with real 
couplings, 
one might be tempted to invoke positivity,
and only the integral over
real fields is manifestly  positive.  However
it is dangerous to invoke reality and positivity 
before taking a thermodynamic limit.   The thermodynamic limit alone does 
most of the 
job of reducing the solution set.  This can seen very explicitly in matrix 
models.
Consider a model of an $N$x$N$ hermitian matrix $M$,
\bg
Z=\int dM e^{-Ntr V(M)}
\label{mam}
\ed
$M$ may be written as $U \Lambda U^{\dagger}$ where $\Lambda$ is diagonal, 
and then if 
the integral over $U$ is performed one gets,
\bg
Z=\int \prod_n d \Lambda_n \Delta^2[\Lambda] e^{ -N\sum_n V(\Lambda_n) }
\label{pro}
\ed
where $\Delta[\Lambda]$ is the Vandermonde determinant 
\bg
\Delta[\Lambda]= \prod_{n<m}(\Lambda_n - \Lambda_m)
\label{vand}
\ed
One can separately choose a contour for each eigenvalue by the condition
$Re V(\Lambda_n) \rightarrow +\infty$ for large $\Lambda_n$.
The boundary condition problem is then most easily understood if 
one first solves the model by the
method of orthogonal polynomials.  The reader is referred elsewhere for 
details \cite{matrix}.
To summarize the method very briefly,  polynomials $P_n(\lambda) = 
\lambda^n + ...$
are defined such that,
\bg
\int d \lambda e^ {- NV(\lambda) }P_n(\lambda)P_m(\lambda) = h_n\delta_{n,m}
\label{ort}
\ed
where the integral is over some permissible complex contour or sum of 
complex contours,
\bg
\int d \lambda = \sum_i a_i \int_{\Gamma_i} d \lambda
\label{sqw}
\ed  
Note that it is assumed that the coefficients $a_i$ of the integral 
contours are 
the same for 
each eigenvalue $\Lambda_n$, therefore the orthogonal polynomial method 
actually does
not include all possible boundary conditions of the Schwinger-Dyson 
equations.
The $P_n$ have the property that 
\bg
\lambda P_n(\lambda) = P_{n+1}(\lambda) + S_n P_n(\lambda) + R_n 
P_{n-1}(\lambda) 
\label{hog}
\ed
The partition function and the Green's functions may be written in terms 
of the 
coefficients $R_n$ and $S_n$
These coefficients satisfy a set of nonlinear recursion relations, known 
as the discrete 
string equations.  
A thermodynamic limit,  or $N\rightarrow\infty$
planar limit,  is obtained if  the $R_n$ and $S_n$  may be written in this 
limit in terms 
of smooth functions of 
$x={n \over N}$ on the interval $[0,1]$.   These 
functions are easily obtained as the local fixed points of the recursion 
relations.
One can then look for the possible contours which give such limiting 
functions, using the 
fact the initial conditions of the discrete string equations, $S_0$ and 
$R_1$, are the
one and two point connected greens functions of the
zero dimensional theory  with action $NV(\lambda)$,  and with 
some choice of a 
sum over contours.
Some initial conditions will be attracted to one of the fixed point 
solutions, leading to 
a thermodynamic limit,  where as others will not.   
In this way it is  easy to see that the solutions associated with a very 
large number 
of boundary conditions  either coalesce or have no thermodynamic limit as 
$N \rightarrow \infty$ 
At certain critical values of the couplings, such as the points about 
which a double scaling
limit may be taken,  the functions of $x$ which one obtains in the planar 
limit become complex 
within the interval [0,1].  This means that as one crosses these critical 
domains, 
many of the 
boundary conditions which give real solutions no longer have a 
thermodynamic limit. 
A large change in the boundary conditions which lead to a thermodynamic 
limit is a general 
phenomenon which occurs as one crosses a phase boundary.  

There are some very exotic boundary 
conditions in matrix
models which lead to a physical thermodynamic limit.  It has been shown 
\cite{chung} 
that in the planar limit the potential $V(\lambda)= {g\over 4}\lambda^4 + 
{\mu\over 2}\lambda^2$ with $\mu < 0$ and $g$ positive
has several solutions with 
varying degrees of symmetry 
breaking,  and that in the double scaling limit these extend to a larger 
number of solutions 
some of which differ perturbatively (beyond first order in the string 
coupling)  and others 
which differ only non-perturbatively.  That one can account for these 
various solutions
with sums over different contour integrations has also been shown 
\cite{davide} \cite{kitaev}.
However it is a highly nontrivial problem to find the boundary conditions 
associated with 
these solutions if one switches on a symmetry breaking term, so that 
$V(\lambda)= {g\over 4}\lambda^4 + 
{\mu\over 2}\lambda^2 + \sigma\lambda$.  The reason for the difficulty is 
as follows.
Consider an arbitrary set of boundary 
conditions (sums over contours)  and consider what happens to the initial 
conditions of the
string equations, $S_0$ and $R_1$,  as $N\rightarrow\infty$.  In this limit
$S_0$ and $R_1$ are determined by the leading term in a saddle point 
expansion.  For $\sigma =0$ there
are two degenerate minima,  and $S_0$ and $R_1$ can approach a continuum 
of possible values,
subject to the constraint $R_1 = {-\mu\over g} - S_0^2$.   However for 
non-zero $\sigma$
there are no degenerate extrema,  and for any set of contours the result 
for $S_0$ and $R_1$
is dominated by the extremum with the lowest potential among those which 
the contours pass through. 
Therefore as $N\rightarrow\infty$ one can only approach a discrete set of 
values rather than
a continuous set: 
$R_1\rightarrow 0$ and $S_0 \rightarrow \lambda_i$ where $\lambda_i$  are 
the three 
extrema of the potential $V(\lambda)$.  These initial conditions lead to a 
smaller solution set than
was obtained in the case where the potential was degenerate.  It appears 
that certain solutions
vanish,  or rather collapse into other solutions,  upon turning on the 
symmetry breaking term.  
Naively that would seem to imply that the solutions which vanish have, at 
$\sigma = 0$, 
singular Greens functions arising from  differentiation of $Z$ with respect 
to $\sigma$.   But
one can readily check that there are no such singularities.
These solutions may be smoothly evolved away
from $\sigma =0$ according to the action principle. Therefore there should 
be some 
set of contours which yields these solutions both at $\sigma = 0$ and at 
some small non-zero
$\sigma$.  We have just seen that the usual choice of a single measure
for every eigenvalue does not work. 
However there are many other boundary conditions of the Schwinger-Dyson or 
loop equations for
which orthogonal polynomial techniques are not appropriate.  For instance 
one may have
staggered boundary conditions with different contours of integration for 
different eigenvalues.   

There are several phenomena which occur in the thermodynamic limit which 
appear to be 
disparate.
One is the collapse of the space of boundary conditions,  due to the fact 
that many 
boundary conditions do not lead to thermodynamic limit.  The other is the 
appearance of 
new non-analyticity and phase boundaries in the coupling constants due to 
the accumulation of
Lee-Yang zeroes \cite{yang}. In 
fact these are not really disparate.  We will give a heuristic argument 
below.   The collapse of the space of boundary conditions means that as 
one varies the couplings,
one must also vary the boundary conditions (contours),  since the set of 
boundary conditions with
a thermodynamic limit depends upon where one is in the space of coupling 
constants.  It is then 
possible that by following a closed path in coupling constant space,  one 
does not return to
the boundary condition with which one started.  This would mean that the 
thermodynamic limit 
has introduced a non-analyticity in the coupling constants,  which in turn 
requires an accumulation 
of Lee-Yang zeroes.  Note that there are simple zero dimensional analogues 
of this.  Consider
the polynomial action $S=\sum_{n=1}^k g_n \phi^n$.  The solutions are 
analytic in in all but the
highest coupling constant $g_k$.  If one follows a 
closed path in the complex plane with any of the lower
coupling constants,  one does not have to change the contour to maintain 
convergence,  and thus
there is only a single Riemann sheet in the lower coupling constants.  If 
however one tunes $g_k$
to zero,  then the space of boundary conditions shrinks.  If one now 
rotates the phase of $g_{k-1}$
by a large amount, then one must also rotate the contour of integration,  
whereas 
the contour could be held fixed
for non-zero $g_k$.  Green's functions  become multiply sheeted in 
$g_{k-1}$ 

The conventional tool for studying the phase structure of theories with a 
local order parameter is
the effective potential.  It is interesting to note the consequences of 
our analysis of
boundary conditions on the form of the effective potential.  In fact,  far 
from the thermodynamic
limit there are many effective potentials.  Consider again the zero 
dimensional $\phi^4$ theory.
In terms of the function $\phi (J) = {\del\over \del J}ln Z(J)$,  the 
Schwinger-Dyson equation is a 
nonlinear second order differential
equation,
\bg
J = g\phi^3 + \mu\phi + g{\del^2\over\del J^2}\phi (J) + 3g\phi (J) 
{\del\over\del J}\phi (J) 
\label{fisj}
\ed
The effective potential $\Gamma(\phi)$ is defined
by the relation $J= {d\Gamma\over d \phi}$.  
Any effective potential can only correspond to a discrete set  of 
solutions $\phi(J)$ out of the full
continuous 2 parameter class.  In fact in zero dimensions $Z(J)$ is 
analytic in $J$,  and
$\phi(J)$ has no branch cuts in $J$, 
so their is only one solution associated with each effective potential.  
Note that order by order
in a loop expansion, $\phi(J)$ is multiply sheeted and the effective 
potential appears
to possess multiple extrema  which correspond to different boundary 
conditions.  This is a spurious
feature of the loop expansion in zero dimensions.
Actually $\phi(J)$ has an infinite tower of 
poles in $J$ rather than branch cuts.  In an appropriate thermodynamic 
limit these poles 
can coalesce,
in which case several boundary conditions become described by the same 
exact effective potential.  
Many effective potentials associated with different boundary conditions 
coalesce in the 
thermodynamic limit,  while effective potentials associated with other 
boundary conditions
do not have a well defined thermodynamic limit.            
\\[24pt]
\leftline{}
\noindent{\bf Conclusion}\\[12pt]
\leftline{}
\indent The phase structure of bosonic field theories and 
matrix models appears to have a very natural interpretation 
in terms of the boundary conditions of Schwinger-Dyson equations. 
There are several interesting open questions of which we will list only a 
few.  
It is not known whether or how these ideas may be generalized to fermionic 
theories.
Another question concerns the relation between the
various boundary conditions and the phases of a theory with a nonlocal 
order parameter.
As yet we have not been able to construct such a relation.

\leftline{}
\leftline{}
\leftline{}
\leftline{}

\noindent{\bf Acknowledgment}\\[12pt]
\leftline{}
This work was done in collaboration with
S. Garcia and G. Guralnik.  A detailed paper on the subject is in 
preparation.
We thank Stephen Hahn for many useful discussions.  The author acknowledges
the support of the Packard Foundation.

\end{document}